\documentclass[12pt]{article}
\usepackage{amsfonts}
\usepackage{amssymb}
\usepackage{graphicx}
\usepackage{color}
\usepackage{amsmath}
\usepackage{amsthm}
\usepackage{fullpage}
\usepackage{natbib}
\usepackage[footnotesize,bf]{caption}

\newcommand{\bs}[1]{\boldsymbol{#1}}

\newcommand{\mr}[1]{\mathrm{#1}}
\newcommand{\bm}[1]{\mathbf{#1}}

\newcommand{\code}[1]{\texttt{#1}}
\newcommand{\e}[1]{{\footnotesize$\times10$}{$^{#1}$}}

\begin{document}

\title{\vspace{-1cm}Estimating Player Contribution in Hockey with Regularized Logistic Regression}

\author{Robert B. Gramacy\\ {\it University of Chicago Booth School of
  Business}\vspace{.3cm}\\
Shane T.~Jensen\\
 {\it The Wharton School,
University of Pennsylvania }\vspace{.3cm}\\
Matt Taddy\\  {\it University of Chicago Booth School of Business } }

\date{\today}

\maketitle

\begin{abstract}
  We present a regularized logistic regression model for evaluating
  player contributions in hockey. The traditional metric for this
  purpose is the plus-minus statistic, which allocates a single unit
  of credit (for or against) to each player on the ice for a goal.
  However, plus-minus scores measure only the marginal effect of
  players, do not account for sample size, and provide a very noisy
  estimate of performance.  We investigate a related regression
  problem: what does each player on the ice contribute, beyond
  aggregate team performance and other factors, to the odds that a
  given goal was scored by {\it their} team?  Due to the large-$p$
  (number of players) and imbalanced design setting of hockey
  analysis, a major part of our contribution is a careful treatment of
  prior shrinkage in model estimation.  We
  showcase two recently developed techniques -- for posterior
  maximization or simulation -- that make such analysis feasible.
  Each approach is accompanied with publicly available software and we
  include the simple commands used in our analysis.  Our results show
  that most players do not stand out as {\it measurably} strong
  (positive or negative) contributors. This allows the stars to really
  shine, reveals diamonds in the rough overlooked by earlier analyses,
  and argues that some of the highest paid players in the league are
  not making contributions worth their expense.

\end{abstract} 

\section{Introduction}
\label{sec:intro}

Player performance in hockey is difficult to quantify due to the
continuity of play, frequent line changes, and the infrequency of
goals.  Historically, the primary measure of individual skater
performance has been the {\it plus-minus} value: the number of goals
scored by a player's team minus the number of goals scored by the
opposing team while that player is on the ice.

More complex measures of player performance have been proposed to take
into account game data beyond goal scoring, such as hits or face-offs.
Examples include the adjusted minus/plus probability approach of
\cite{schlocwel10} and indices such as Corsi and DeltaSOT, as reviewed
by \cite{vol10}.  Unfortunately, analysts do not generally agree on
the relative importance of the added information.  While it is
possible to statistically {\it infer} additional variable effects in a
probability model for team performance \cite[]{ThoVenJen12} our
experience is that, in the low-scoring world of hockey, such
high-dimensional estimation relies heavily upon model assumptions that
are difficult to validate. As a result, complex scores provide an
interesting new source of commentary but have yet to be adopted as
consensus performance metrics or as a basis for decision making.

Due to its simplicity, the plus-minus remains the most popular measure
of player performance.  It has been logged for the past fifty years
and is easy to calculate from the current resolution of available game
data, which consists of the identities of each player on the ice at
any time point of the game as well as the times when goals were
scored.  However, a key weakness is that the plus-minus for each
player does not just depend on their individual ability but also on
other factors, most obviously the abilities of teammates and
opponents.

In statistical terms, plus-minus is a \emph{marginal
  effect}: it is an aggregate measure that averages over the
contributions of opponents and teammates.  Since the quality of the
pool of teammates and opponents that each player is matched with
on-ice can vary dramatically, the marginal plus-minus for individual
players are inherently polluted.  Another disadvantage is that
plus-minus does not control for sample size, such that players with
limited ice-time will have high variance scores that soar or sink
depending on a few chance plays.

A better measure of performance would be the \emph{partial
    effect} of each player, having controlled for the contributions
of teammates, opponents and possibly other variables.  To this end, we
propose a logistic regression model to estimate the credit or blame
that should be apportioned to each player when a goal is scored.  In
keeping with the spirit of plus-minus (and using the same publicly
available data), we focus on the list of players on the ice for each
goal as our basic unit of analysis.  Briefly, denote by $q_{i}$ the
probability that a given goal `$i$' was scored by the home team ({\it
  home} and {\it away} are just organizational devices; results are
unchanged upon modeling $\mr{p}(away)$ instead).  Then
\begin{equation}\label{logodds}
  \log\left(\frac{q_i}{1-q_i}\right) = \alpha_i + \beta_{h_{i1}} +
  \ldots + \beta_{h_{i6}} - \beta_{a_{i1}} \ldots
- \beta_{a_{i6}},
\end{equation}
where $\bs{\beta} = [\beta_1 \cdots \beta_{n_p}]'$ is the vector of
{\it partial plus-minus effects} for each of $n_p$ players in our
sample, with $\{h_{i1} \ldots h_{i6}\}$ and $\{a_{i1} \ldots a_{i6}\}$
being the indices on $\bs{\beta}$ corresponding to home-team ($h$) and
away-team ($a$) players on the ice for goal $i$.\footnote{Note that we
  include goalies in our analysis.} The intercept term $\alpha_i$ can
depend upon additional covariates, e.g. the model we present in
Section~\ref{sec:models} incorporates team indicators into $\alpha_i$.

In almost all regressions, one is susceptible to the twin problems of
{\it over-fit}, where parameters are optimized to statistical noise
rather than fit to the relationship of interest, and {\it
  multicollinearity}, where groups of covariates are correlated with
each other making it difficult to identify individual effects.  These
issues are especially prominent in analysis of hockey, with a high
dimensional covariate set (around 1500 players in our sample) and a
very imbalanced {\it experiment design} -- due to use of player lines,
wherein groups of 2-3 players are consistently on ice together at the
same time, the data contain many clusters of individuals who are
seldom observed apart.  

A crucial contribution of our paper is a careful treatment of the partial player effects $\bs{\beta}$ that helps alleviate both problems.  As outlined in Section
\ref{sec:models}, a {\it prior distribution} with its mode at the
origin is placed on each $\beta_j$; this adds a {\it penalty} -- e.g.,
$\lambda_j\beta_j^2$ or $\lambda_j|\beta_j|$ -- on the likelihood
function and shrinks estimates of this coefficient towards zero.   
We use the term  {\it regularization} to refer to the shrinkage of regression coefficients towards zero that is induced by our prior distribution.   

Building on new developments in regularized estimation for regression
with binary response (such as our home-vs-away outcome), we leverage
machinery that has only very recently become available for data sets
of the size encountered in our analysis.  In particular, we detail
penalized likelihood maximization for fast robust estimates of player
contribution, as well as Bayesian simulation for exploring joint
uncertainty in multiple player effects, allowing for comparisons
between groups of players which would not have been possible with
earlier methodology.  In both cases, inference proceeds through simple
commands to newly developed packages for the open source {\sf R}
analysis software \citep{cranR}.  The resulting player effects are
easy to interpret, and our hope is that readers will experiment
with our models to feed a discussion on alternative
plus-minus metrics.

The remainder of the paper is outlined as follows. Section
\ref{sec:background} provides an overview of previous attempts at a
partial player affect.  These avoid full-scale logistic regression
which, until very recently, would not have been computationally
feasible.  Section \ref{sec:reglogit} details our data and general
regression model.  Section \ref{sec:pointe} presents point estimates
of player effects, comparing results both with and without controlling
for teams and under a range of levels of prior regularization.  These regularized point estimates can be used for variable selection, since only a subset of effects will be non-zero.   Section \ref{sec:decision} then describes a full Bayesian analysis of
the joint uncertainty about players, and illustrates how such
information can be used by coaches and general managers to make
important personnel decisions. 
The paper concludes in Section \ref{sec:discuss} with a discussion,
and an appendix which contains details of our estimation algorithms
and the entertainment of a full interaction model.

\subsection{Background on Adjusted Plus-Minus}
\label{sec:background}

The strategy of conditional estimation for player ability is not new
to sports analysis.  For example, \citet{awa09} advocates a simple
adjustment to hockey plus-minus that controls for team strength by
subtracting team-average plus-minus.  Basketball analysts have been
active with conditional performance models, including the linear
regressions employed by \citet{ros04} and \citet{ilabar08}.  Due to
frequent scoring and variability in the combination of players on the
floor, estimation of partial effects is generally easier in basketball than in
the low scoring and imbalanced design setting of hockey.

For hockey, \citet{mac10} proposes analysis of the relationship
between players and goals through regression models similar to those
used in basketball.  He tracks the length of time on ice and goals
scored in each of around 8\e{5} ``shifts'' -- unique combinations of
players -- to build a goals-per-hour response that is regressed onto
player effect variables analogous to our $\bs{\beta}$ in
(\ref{logodds}).  While Macdonald's work is related to ours, as both
are regressing scoring onto player presence, we outline several
distinctions between the approaches which are helpful in understanding
the motivation behind our modeling.

The use of shift-time (regardless of whether a goal was scored)
introduces extra information but also leads to questions on data
quality and model specification.  For example, we find a median actual
shift length of eight seconds, such that the recorded total
time-on-ice for unique player combinations is built from pieces that
may be too small for play to develop or be assessed.  The average
goals-per-shift is around 0.02 such that less than 2\% of the sampled
response is nonzero.  This calls into doubt the assumption of
approximate normality upon which Macdonald's standard error estimates
are based.  Moreover, although there are more observations in the
sample than there are covariates, a vast majority of scoreless shifts
implies that least-squares estimation is dominated by a few scoring
shifts.  Even with a more balanced sample, estimation on this
dimension is likely overfit without regularization. Indeed, Macdonald
reports only top skaters among those with at least 700 minutes on ice;
we suspect that players with very little ice time dominate
the full list of effects.

\section{Data and Model}
\label{sec:reglogit}

This section details our data, the full regression
model, and prior specification.

\subsection{Data}
\label{sec:data}

The data, downloaded from {\tt www.nhl.com}, comprise of information
about the teams playing (with home/away indicators), and the players
on ice (including goalies) for every even strength goal during the
four regular seasons of 2007--2008 through 2010--2011.  There were
$n_p = 1467$ players involved in $n_g = 18154$ goals.  In keeping with
the canonical definition of plus-minus, we do not consider
power-play/shorthanded or overtime goals; however, our framework is
easily extended to handle such data. 

This article focuses on the 4-season data aggregate and treats player
ability as constant over this range.  The larger data window allows
for better effect measurement, especially due to the increased
variation in on-ice configuration and player movement between teams.
However, different aggregations will be desirable for different
situations and one can apply our methods to any subset.  Given a data
window that is too short for measurement of individual ability, our
sparse estimation methods will simply set all player effects to zero.

\begin{figure}[ht!]
\centering
\includegraphics[scale=0.45]{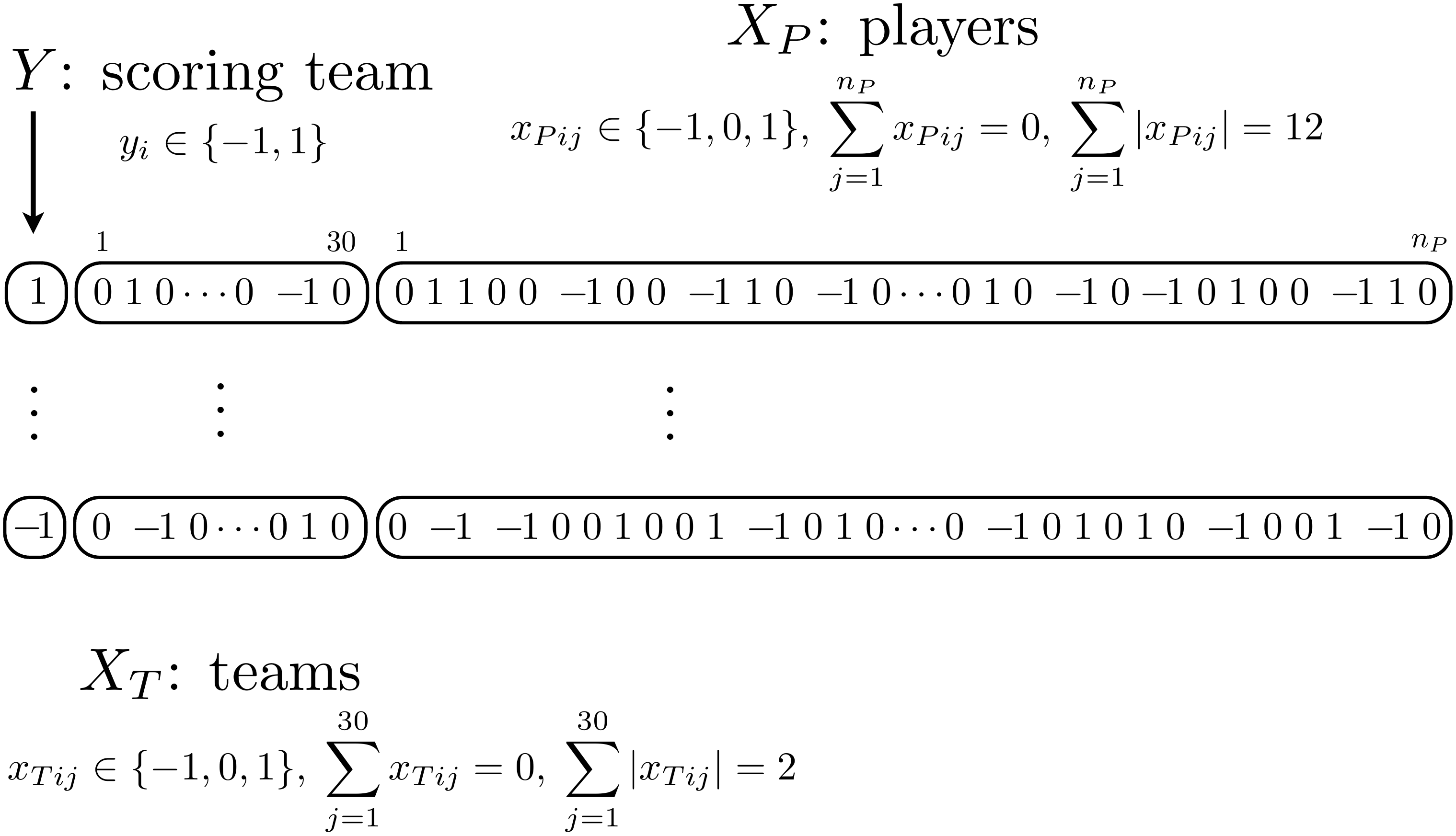}
\caption{A diagram of the design matrix and two example rows.  Two
  goals are shown under the same configuration of teams and players,
  except that the home team has scored in the first case and the
  visiting team in the second (so that the two rows have opposite
  parity).  Exactly two teams have nonzero entries and exactly twelve
  players (six home, six away) are nonzero in each row.  }
\label{f:data}
\end{figure}

The data are arranged into a response vector $Y$ and a design matrix
$X$ comprising of two parts, $X_T$ and $X_P$, for indicator variables
corresponding to team and player identities respectively.  For each
goal $i$ the response vector contains $y_{i} =1$ for goals scored by
the home team and $y_i=-1$ for away team goals.  The corresponding
$i^\mathrm{th}$ row of $X$ indicates the teams playing and the players
on the ice when that goal was scored, with $x_{Tij}$ equal to 1 for
the home team and $x_{Pij}$ equal to 1 for each home player on the
ice, and $x_{Tij}$ equal to $-1$ for the away team and $x_{Pij}$ equal
to $-1$ for each away player on the ice.  All other $x_{Tij}$ and
$x_{Pij}$ indicators are equal to zero.  Figure \ref{f:data}
illustrates this data structure with two example rows.

Note that the design matrix is extremely sparse: overall dimensions
are $n_g \times (n_p + 30) = 18154 \times 1497$, but every row has
1485 zeros for more than 99\% sparsity.  As already noted,
the design is also highly imbalanced:  only about 27K of the greater than
one million possible player pairs are actually observed on the ice for a
goal.

\subsection{Logistic Likelihood Model}
\label{sec:models}

From the data definition of \ref{sec:data}, we can reformulate our
logistic regression equation of (\ref{logodds}) as 
\begin{equation}\label{fullmodel}
 \log\left(\frac{q_i}{1-q_i}\right) = \bm{x}_{Ti}'\bs{\alpha} + \bm{x}_{Pi}'\bs{\beta}
\end{equation}
where $q_i = \mr{p}(y_i = 1)$, $\bm{x}_{Pi} = [x_{Pi1}\cdots
x_{Pin_p}]'$ is the vector corresponding to the $i^{th}$ row of $X_P$,
$\bm{x}_{Ti}$ is similarly the $i^{th}$ row of $X_T$, $\bs{\alpha}$ is
the length-30 vector of team effects and
$\bs{\beta}$ is the length-$n_p$ vector of player effects.

The likelihood model in (\ref{fullmodel}) is easily extended to
incorporate additional conditioning information or more flexible
player-effect specifications.  For example, $\bm{x}_{Ti}$ and
$\bs{\alpha}$ could be lengthened to account for special teams effects
(e.g., including variables to indicate penalties or other
non-five-on-five situations) or potential sources of bias (e.g.,
referee indicators).  Moreover, $\bm{x}_{Pi}$ and $\bs{\beta}$ could
be doubled in length to include distinct offensive and defensive
player effects, as in \cite{ilabar08}.  We focus on the current
formulation to stay true to the spirit of the standard plus-minus
metric, although we do investigate a model with pairwise interactions
between players in Appendix~\ref{interactionresults}.

\subsection{Bayesian Approach and Prior Regularization}

We impose shrinkage on our regression coefficients by taking a
Bayesian approach that combines our logistic likelihood model
(\ref{fullmodel}) with a prior distribution centered at zero.  The
general Bayesian approach consists of a likelihood $p(y | \theta)$
model that specifies the observed data as a function of unknown
parameters $\theta$ and a prior distribution $p(\theta)$ for them.  In
this particular application, the unknown parameters $\theta$ are the
team partial effects $\bs{\alpha}$ and player partial effects
$\bs{\beta}$.  Inference in the Bayesian approach is based the
posterior distribution $p(\theta | y) \propto p(y | \theta) \times
p(\theta)$.  In a full Bayesian analysis, the entire posterior
distribution is estimated, usually via simulation-based methods such
as Markov chain Monte Carlo.  Alternatively, we can restrict ourselves
to the maximum {\it a posteriori} (MAP) estimates of the unknown
parameters.

Imposing a prior distribution on the regression coefficients of our
logistic model (\ref{fullmodel}) is necessary to guard from overfit
and provide stable estimates of individual player effects.  To
emphasize this point, consider an attempt at maximum likelihood
estimation for the simplest non-team-effect model
($\bm{x}_{Ti}'\bs{\alpha}$ replaced by shared $\alpha$).  Fitting the
standard logistic regression model\footnote{ Fitted in {\sf R} using
  the command: {\tt fit <- glm(goals $\sim$ XP, family="binomial")}}
yields a half hour wait and several numerical errors.  Forward
step-wise regression\footnote{We used forward step-wise regression
  with the Bayes information criterion (BIC)} is not a solution: it
takes hours to converge on a model with only three significant
players.

Using prior distributions $\pi(\alpha_j)$ and $\pi(\beta_j)$ for each
model parameter adds stability to the fitted model by shrinking each
coefficient towards a central value of zero.  From the perspective of
point estimation, placing a prior distribution on $\beta_j$ (or
$\alpha_j$ equivalently) centered at zero is equivalent to adding a
penalty term for $\beta_j \neq 0$ in the objective function that is
being optimized to give us point estimates for each $\beta_j$.

Different types of prior distributions correspond to different penalty
functions on $\beta_j \neq 0$.  One common strategy uses a normal
prior distribution (centered at zero) on each $\beta_j$ which
corresponds to a L2 penalty ($\lambda_j \beta_j^2$) in the objective
function.  Thus, the MAP estimates from a Bayesian regression model
with a normal prior distribution are equivalent to the estimates from
ridge regression \citep{HoeKen70}.  A normal prior distribution (L2
penalization) is used under an assumption that every covariate has a
limited effect on the response, i.e. elements of $\bs{\beta}$ are
non-zero but small.

Another popular strategy uses a Laplace prior distribution on each
$\beta_j$ which corresponds to a L1 penalty ($\lambda_j |\beta_j|$) in
the objective function.  Thus, the MAP estimates from a Bayesian
regression model with a Laplace prior distribution are equivalent to
the estimates from {\it lasso} regression \citep{Tib96}.  A Laplace
prior distribution yields a penalized point estimate of exactly
$\beta_j=0$ in the absence of strong evidence, and non-zero $\beta_j$
for only a subset of significant variables.  In this way, using an L1
penalty naturally permits {\it variable selection} where only a subset
of covariates are selected as having substantive predictive effects on
the outcome variable.
  
We favor an L1 penalty/Laplace prior for player effects $\bs{\beta}$
because it permits {\it variable selection}: the identification of
players that stand out as having truly substantive effect.
We also employ a L2 penalty/normal prior for coefficients on
``nuisance'' variables, such as the team effects $\bs{\alpha}$.  L1
and L2 penalties are fairly standard choices for sparse and dense
models respectively: L2 has a long history in Tikhonov regularization
and ridge regression, while L1 has a shorter history but wide usage in
Lasso estimation.  The focus on influential players yielded by
coefficient sparsity is key to our analysis and exposition, while
dense estimation is a more conservative choice for the controlling
variables since it does not assume their effect can be represented in
a lower dimensional subspace (results are practically unchanged if one
uses L1 regularization for the team effects).

For completeness, we summarize here that this combination of L1 and L2
penalty terms is built into our Bayesian model by imposing the
following prior densities on our regression coefficients,
\begin{equation}\label{eqprior}
\pi(\bs{\alpha},\bs{\beta}) =
\prod_{t=1}^{30}\mr{N}(\alpha_t; 0,\sigma_t^2)
\prod_{j=1}^{n_p}\mr{Laplace}(\beta_j; \lambda_j).
\end{equation}
We begin our analysis by investigating the maximum {\it a posteriori}
(MAP) estimates of $\bs{\alpha}$ and $\bs{\beta}$ in
Section~\ref{sec:pointe}.  We then explore the full posterior
distribution of our regularized logistic regression model in Section
\ref{sec:decision}.  Specification of prior parameters $\sigma_t$ and
$\lambda_j$ dictates the amount of penalization imposed on estimates,
and each analysis section outlines its approach and sensitivity to
this choice.  In particular, our MAP estimation jointly optimizes over
both the coefficients and their penalty, while the fully Bayesian
analysis averages player effects over possible values of a single
shared penalty.  Full estimation algorithms and software description
are in Appendix~\ref{sec:methods}.

\section{Point Estimates of Player Contribution}
\label{sec:pointe}

This section presents MAP point estimates for $\bs{\beta}$, the player
contribution partial effects, under the regression in
(\ref{fullmodel}) with priors in (\ref{eqprior}).  Team-effect
prior variances are fixed at $\sigma_t = 1$, giving a standard normal prior 
specification.  Due to the large amount of likelihood information on
team effects, our results are largely insensitive to the value of  $\sigma_t$.  

The Laplace prior parameters $\lambda_j$ for our player effects
require more care.  We place an additional {\it hyperprior}
distribution on the $\lambda_j$ parameters so that the data can help
us infer these penalty parameters along with their coefficients.  The term {\it hyperprior} is used for parameters such as $\lambda_j$ that are themselves involved in the prior distributions of our main parameters of interest (player effects $\bs{\beta}$).   A model with several levels of unknown parameters (and prior distributions) is often called a {\it Bayesian hierarchical model}. 

Following \citet{taddy:2012}, independent conjugate gamma prior
distributions are assumed for each $\lambda_j$ with
$\mathrm{var}[\lambda_j] = 2 \times \mathrm{E}[\lambda_j]$.
Throughout Section \ref{sec:pointe}.1 we use $\mathrm{E}[\lambda_j] =
15$ which was smallest penalty we could manage while eliminating large
nonzero $\beta_j$ for players with very little ice time.  Note also
that posterior samples for a {\it shared} L1 penalty $\lambda$ are
obtained as a byproduct of the full Bayesian analysis in
Section~\ref{sec:decision}.  The inferred posterior mean
$\mathrm{E}[\lambda]$ is again around 15, further justifying a MAP
estimation centered on this value.  

Details of our model implementation are given in
Appendix~\ref{sec:methods}.  Section \ref{sec:pmcomp} offers a
comparison to traditional plus-minus, and a sensitivity analysis for
the prior parametrization is given in Section~\ref{sec:sens}.  We
conclude by augmenting with salary information in Section
\ref{sec:value} in order to comment on player value-for-money.

\subsection{MAP Estimation of Partial Player Effects}

\begin{figure}[t]
\centering
\includegraphics[scale=0.48,trim=50 29 35 10]{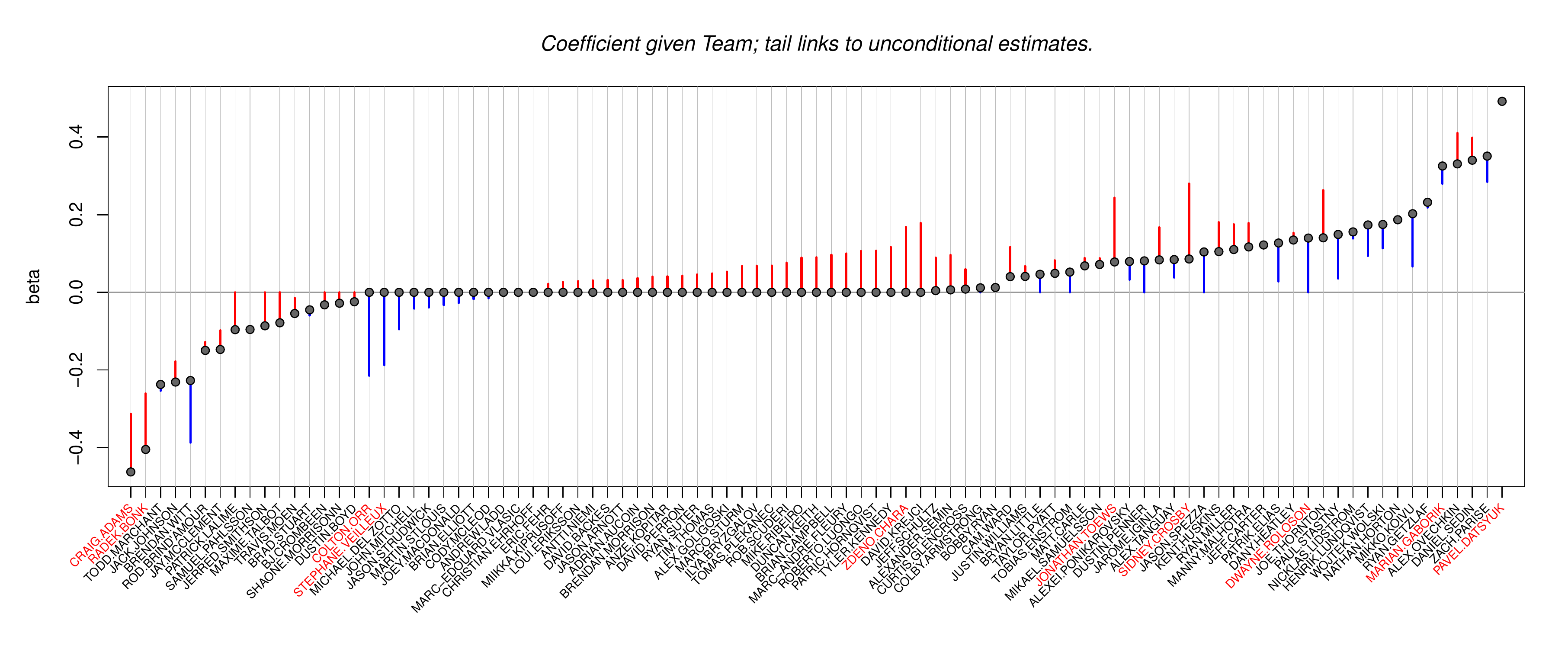}
\caption{Comparing main effects for players in the team-augmented model
  (dots), to the player-only model.  The lines point to the
  unconditional (player-only) estimates.  The coefficients have been
  ordered by the dots.  Players discussed in the text have their names colored in red.  Players with coefficients estimated as zero under both
  models are not shown.}
\label{f:main}
\end{figure}

We consider two models for conditional player effect estimation: the
full team-player model of (\ref{fullmodel}) and a player-only model,
where $\bm{x}_{Ti}'\bs{\alpha}$ is replaced by a single shared
parameter $\alpha$.  Figure \ref{f:main} shows the main effects
obtained for the team--player model under MAP estimation (dots), and
compares to those from a player-only model (connecting lines).  Only
players with non-zero effects in either model are shown.  The $x$-axis
orders the players by their estimates in the team--player model,
expressing a marginal ordering on player ability.

Observe that incorporating team effects causes many more player
effects to be zeroed out, with many players' stand-out performance
being absorbed by their respective teams (red lines tracking to zero).
Note that ability estimates for players who frequently share the ice
will generally be negatively correlated: if players are always on the
ice together when a goal is scored, an increase in the estimated
effect of one player must be accompanied by a decrease in the effect
of the other player.  In MAP estimation under a sparsity-inducing L1
penalty, this will often manifest itself by forcing one player’s
effect being estimated at zero.

Perhaps the most surprising result is that Sidney
Crosby, considered by many to be the best player in the NHL, has a
contribution that drops after accounting for his team (although he
still stands out).  Jonathan Toews' and Zdeno Chara's effects show
similar behavior, the latter having no player--team effect.  As all
three players captain their respective (consistently competitive)
teams, we should perhaps not be surprised that team success is so
tightly coupled to player success in these cases.

An exception is Pavel Datsyuk, who stands out as the leagues very
best, having a coefficient that is unmoved even after considering the
strong team effect of his Red Wings.  There are also a few players,
such as Dwayne Roloson, who shine despite their team.  Roloson has a
strong positive effect in the player--team model but a null one in the
player-only model.  We will revisit this particular result in
Section~\ref{sec:pmcomp}.

On the negative side, Colton Orr and Stephane Veilleux seem to shoulder
undue blame for their team's poor performance.  Orr, recognized as an
enforcer when on the ice, may not have been used effectively by his
coaching staff.  Veilleux played alongside Gaborik on the Wild in the
late 2000s, and both players get a positive bump in the player--team
model.  Finally, Craig Adams and Radek Bonk stand out as poor
performers in both models.

\subsection{Comparison to traditional plus-minus}
\label{sec:pmcomp}

Our variable selection approach provides a rich but compact summary of
player performance.  In our team-player model with an L1 penalty, the
vast majority of players obtain a ``mediocre'' $\hat{\beta}_j = 0$ and
our focus can be narrowed to those who have a significant partial
effect on scoring.  In contrast, plus-minus assigns a non-zero number
for most players without any reference to statistical significance.
Thus the most obvious departure from traditional plus-minus is that
far fewer players are distinguishable from their team-average under
our performance metric.

From the model equation in (\ref{logodds}), nonzero partial player
estimates are an additive effect on the log odds that, {\it given a
  goal has been scored}, it is a goal for that player's team
(controlling for team and the other players).  In other words,
$e^{\beta_j}$ is a multiplier on the for-vs-against odds for every
goal where player $j$ is on the ice, so that our parameters $\beta$
relate multiplicatively to the {\it ratio} of for-vs-against, while
traditional plus-minus is calculated as the {\it difference} between
for-vs-against goals.  Moreover, our partial effects measure
deviations in performance from the team average so that a player on a
very good team needs to be even better than his teammates to gain a
positive $\beta$, while an average player on a good team has an
impressive plus-minus measure just by keeping up.

\begin{figure}[ht!]
\centering
\begin{minipage}{7.5cm}
\includegraphics[scale=0.5,trim=0 10 35 10]{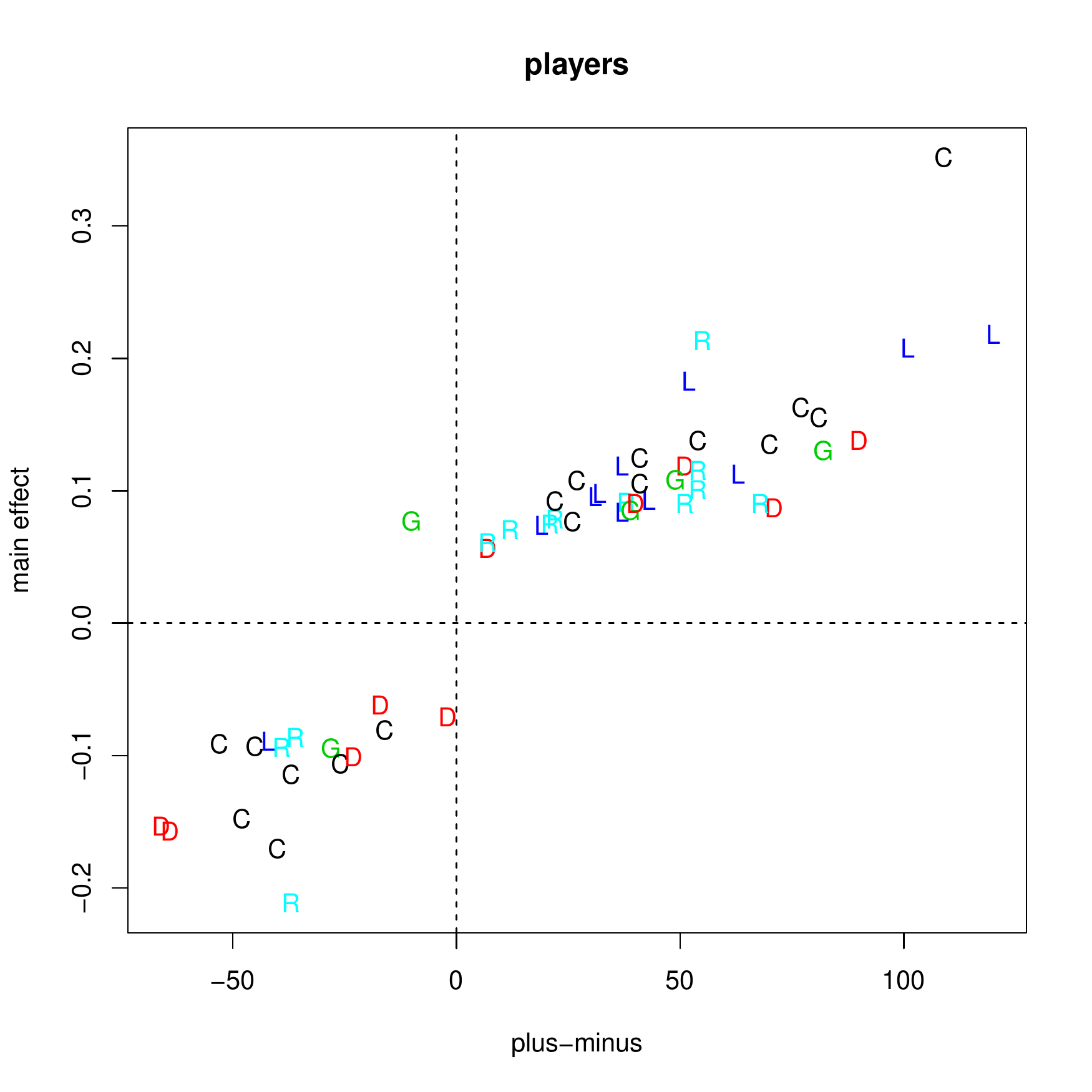}
\end{minipage}
\hspace{2cm}
{\footnotesize 
\begin{tabular}{r|rr}
  team    &     MAP   & $+/-$ \\
\hline
T.B  & $-0.165$ & $-159$ \\
NYI & $-0.126$ & $-143$ \\
EDM & $-0.077$ &  $-103$ \\
ATL & $-0.045$ &  $ -80$ \\
COL & $-0.025$ &  $-60$\\
OTT & $-0.020$ & $ -56$\\
MIN & $-0.019$ &  $-52$\\
TOR & $-0.017$ & $ -70$ \\
NSH  & $0.019$  &  $9$ \\
S.J  & $0.045$  & $ 93$ \\
BOS  & $0.103$ & $109$ \\
CHI  & $0.106$ & $99$ \\
PIT  & $0.112$ & $114$\\
\end{tabular}}
\caption{{\em Left:} Comparing plus-minus, aggregated over the four
  seasons considered in our analysis, to the MAP partial
  effects $\hat{\beta}$. Plot symbols give positional information: C = center, L = left wing, R = right wing, D = defense, and G = goalie.  {\em Right:}
  Comparing team partial effects $\hat{\alpha}$ to their plus-minus values.  }
\label{f:playerpm}
\end{figure}

Despite these differences, both metrics are an attempt to quantify
player contribution.  Figure \ref{f:playerpm} compares our MAP
estimates of player partial effects from the team-player model to the
traditional plus-minus aggregated over our four seasons of data. The
left-hand plot shows the player estimates, labeled by positional
information.  

  
Discrepancies between the two metrics in Figure \ref{f:playerpm} are
informative.  One player, Dwayne Roloson, has a plus-minus whose sign
disagrees with that on his $\hat{\beta}_j$.  We also noted
earlier that Roloson has a partial player effect that is pulled up in the
team--player model compared to the player-only model.  For an
explanation, we can examine the $\hat{\alpha}_j$ coefficients of the
teams which appear in the table on the right-panel of
Figure~\ref{f:playerpm}.  Each of the four teams Roloson played for
(T.B, NYI, EDM, and MIN) all have significantly negative $\hat{\alpha}_j$
values.  Apparently Roloson was a quantifiable star on a string of
poorly performing teams. Our model reasonably attributes many of the
goals counting against him in his traditional plus-minus as counting
against his team as a whole.

Another observation from Figure~\ref{f:playerpm} is that our model
estimates disagree with traditional plus-minus about who is the best
player in hockey.  Alex Ovechkin is the player with the largest
plus-minus value, although there are nearly a dozen other players with
plus-minus values are within ten percent of his.  In contrast, Pavel
Datsyuk is the best player according to our partial player effects by
a huge margin: his posterior odds of contributing to a goal for his
team are nearly 50\% larger than the next best players (Ovechkin and
Gaborik).

The second best player in hockey by traditional plus-minus is Roberto
Luongo.  However, from Figure \ref{f:main}, we see that Luongo's
partial player estimate is $\hat{\beta}_j = 0$.  In the context of a
goalie, the implication is that his play is not significantly
different from that of his back-ups on the Vancouver Canucks.  This
suggests that undue blame and credit may have been placed on Luongo
for both regular season successes and postseason collapses.  At the
other end, observe that the best ranked player with a negative partial
player estimate (Michael Del Zotto) has a nearly-zero plus-minus
value.

\subsection{Prior sensitivity analysis}
\label{sec:sens}

\begin{figure}[t]
\centering
\includegraphics[width=6.5in]{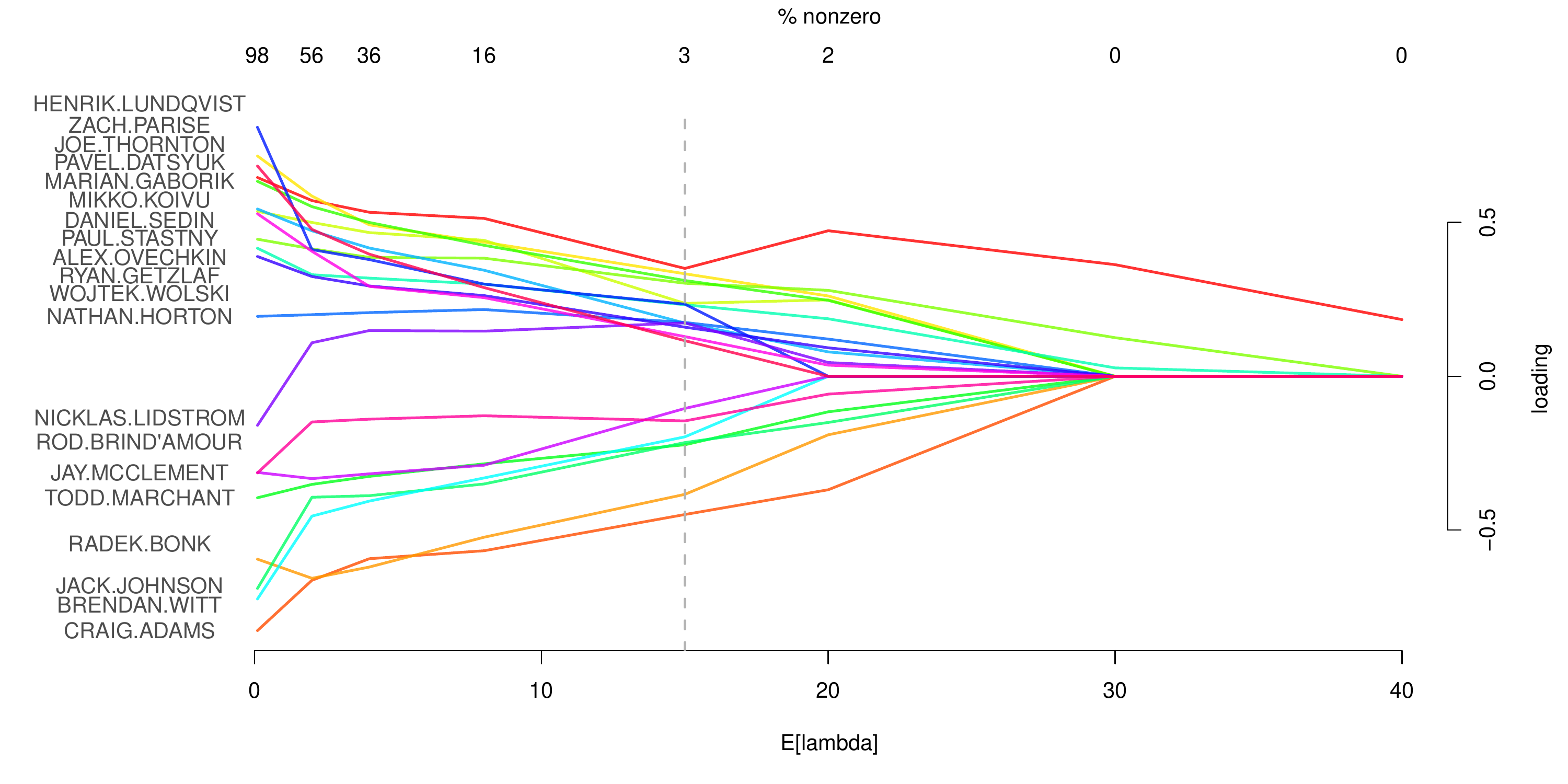}
\caption{Coefficient estimates for a subset of players (chosen from
  all players with nonzero coefficients at $\mr{E}[\lambda_j] = 15$,
  our specification in Sections \ref{sec:pointe}.1-2).  The expected
  L1 penalty is shown along the bottom, with corresponding \% of
  estimated $\beta_j \neq 0$ along the top and coefficient value on
  the right. }
\label{f:sens}
\end{figure}

MAP estimates for each $\beta_j$ are sensitive to $\mr{E}[\lambda_j]$,
the expected L1 penalty for each coefficient.  This relationship is
illustrated in Figure \ref{f:sens}, which shows estimated coefficients
under increasing penalty for some players with large effects in
Sections \ref{sec:pointe}.1-2.  We see that as the expected value of
the penalty term grows, the number of non-zero players (top X-axis of
Figure \ref{f:sens}) decreases substantially, which illustrates the
variable selection aspect of our model.  The magnitudes of the MAP
estimates of partial player effects $\beta_j$ are also shrunk towards
zero with larger penalties.

The far left values have a very low $\mr{E}[\lambda_j] = 1/10$ and
non-zero $\hat{\beta}_j$ for 98\% of players.  At this extreme, there are
three plotted players with stronger effect than Datsyuk.  Only Zach
Parise is more effective than Datsyuk at $\mr{E}[\lambda_j] = 2$,
which leads to 56\% of players with non-zero effects.  At
$\mr{E}[\lambda_j] = 4$, Datsyuk is the top player and 36\% of players
have non-zero effects.  Datsyuk remains the best player at very high
penalization levels, until he is the only measurable contributor in
the league.

More dramatic changes can be found in the (un-plotted) estimates for
players with relatively low ice-time. As an example, Michel Ouellet is
among the top estimated performers in the league for
$\mr{E}[\lambda_j] < 10$, but he jumps to a zero $\hat{\beta}_j$ under
higher penalties.  Given this sensitivity, it is worth revisiting
hyperprior specification.  Although we have chosen $\mr{E}[\lambda_j]$
with help from results of Section \ref{sec:decision}, this value was
also only slightly higher than where non-star players (e.g., Ouellet)
drop out of the top $\hat{\beta}_j$ rankings.  In the absence of reliable
out-of-sample testing, one pragmatic option is to increase penalty
until the results become unrealistic.  Similarly, one can interpret
estimates conditional on the number of nonzero coefficients, and
consider the $\hat{\beta}_j$ as performance measured under a given level of
sparsity.  A better approach, however, is to nearly eliminate
sensitivity to the prior by averaging over penalty uncertainty, as we
do in Section~\ref{sec:decision}.

\subsection{Value for money}
\label{sec:value}

\begin{figure}[ht!]
\includegraphics[scale=0.5]{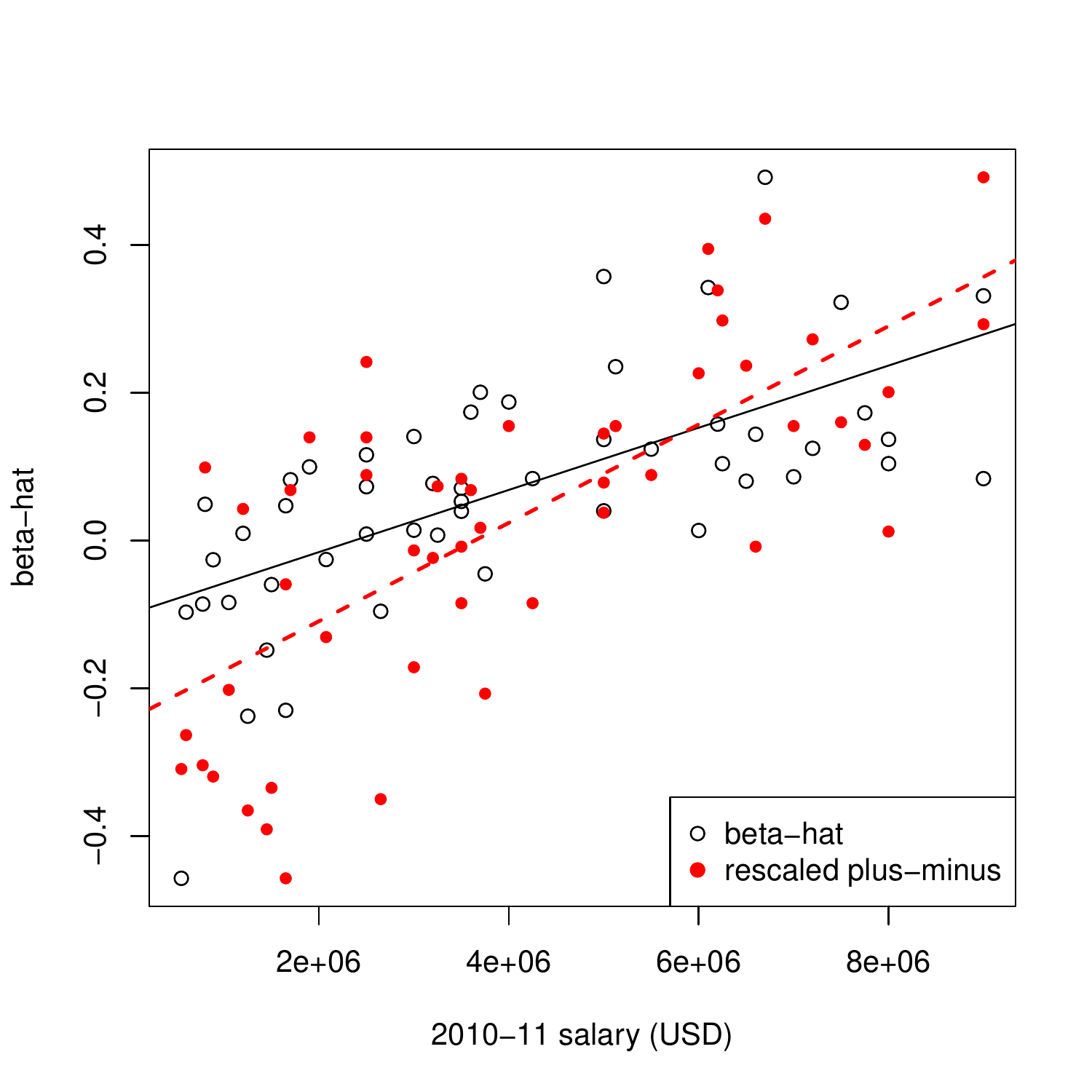}
\hfill
\includegraphics[scale=0.5]{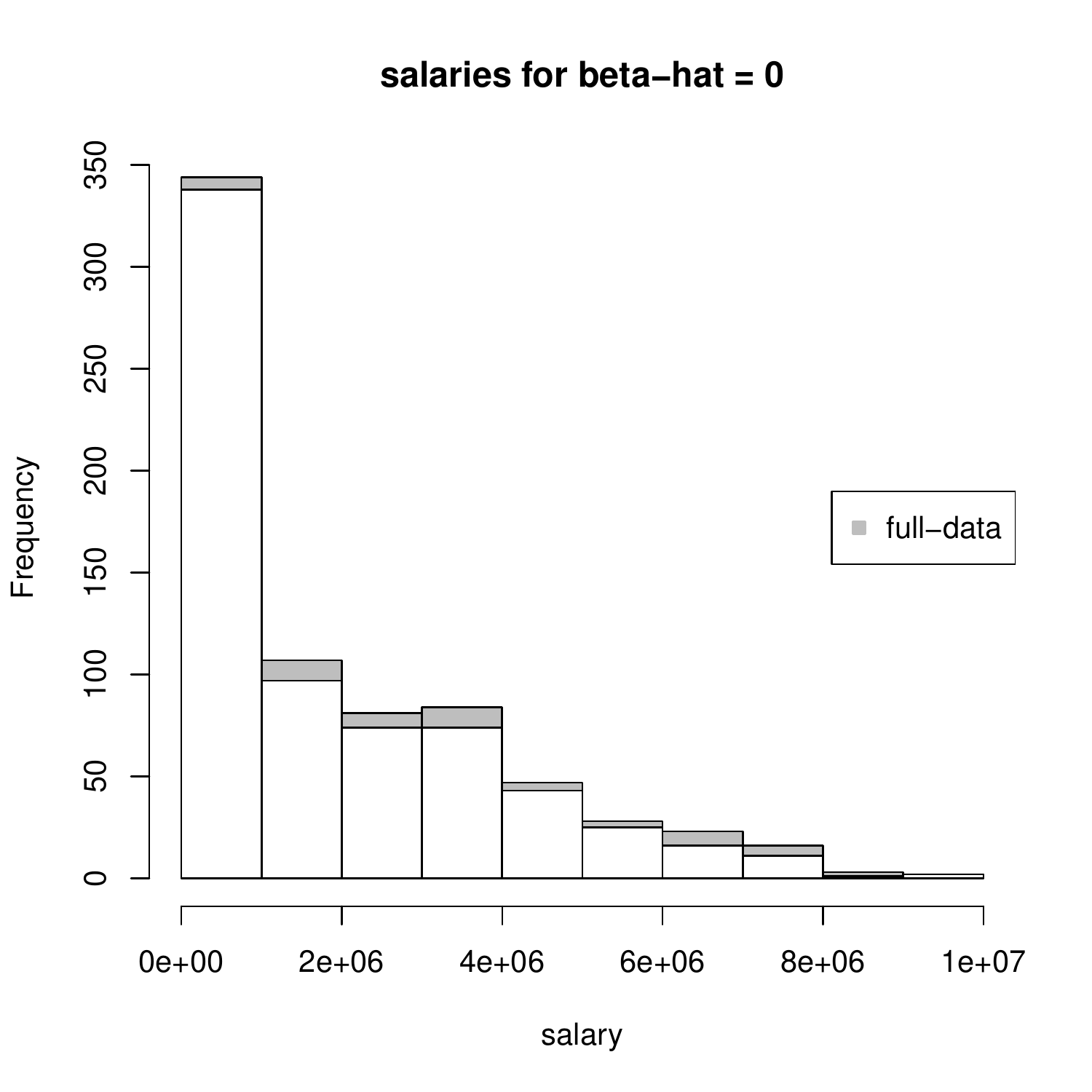}
\caption{The left plot shows non-zero MAP $\hat{\bs{\beta}}$
  estimates versus 2010-11 salary, augmented with rescaled plus-minus
  points for comparison.  Ordinary least squares fits are added to aid
  in visualization.  The right plot shows the histogram of
  2010-11 salaries for players with $\hat{\beta}_j = 0$, extending to the
  full set in gray.}
\label{f:bhatvsal}
\end{figure}

In the left of Figure \ref{f:bhatvsal}, we plot MAP $\hat{\bs{\beta}}$
estimates from our model versus 2010-11 salaries for the non-zero
coefficients.  Plus-minus points (rescaled to fall into the same
range) have also been added to the left plot.

The lines overlaid on the left plot are ordinary least squares fits
for each metric; a hypothesis test for the interaction coefficient
reveals that indeed the two lines differ (at the 5\% level; $p =
0.026$).  Since the $\beta_j$ relate to performance on a log scale, as
is typically assumed for salary, we are not surprised to see a linear
relationship between salary and our player effects.  The fact that the
standard errors for the $\hat{\bs{\beta}}$ fit (0.1226) is less than the
plus-minus fit (0.1605) with the same design inputs suggests that our
model-based player effects $\hat{\bs{\beta}}$ have greater correlation to
player salary.  Teams appear to be compensating their players using
strategies that are more in line with our partial player effects
($\bs{\beta}$) than with traditional plus-minus.

One reason why our model estimates have a lesser slope with salary is
that fewer players are estimated to have a substantial (non-zero)
negative contribution by our model compared to traditional plus-minus.
Despite the decent correlation between our model estimates and player
salary, it is also clear that there are some mis-priced players.  The
best player according to our model, Pavel Datsyuk, probably deserves a
raise whereas Sidney Crosby may be somewhat over-priced.  Alex
Ovechkin seems to be correctly priced in the sense that he is close to
the fitted line between his model estimate and his salary.

In the right of Figure \ref{f:bhatvsal}, we give the salary
distribution for players that were estimated to have zero player
effects.  Clearly, there are many players with high salaries but which
are estimated by our model to not have a substantial player effect.
Specifically, the top ten salaries for players with $\hat{\beta}_j =
0$ are Chris Pronger (\$7.6M), Henrik Zetterberg (\$7.75M), Brad
Richards (\$7.8M), Marian Hossa (\$7.9M), Chris Drury (\$8M), Scott
Gomez (\$8M), Duncan Keith (\$9M), Evgeni Malkin (\$10M), and Vincent
Lecavalier (\$10M).

The gray extensions to the histogram indicate the full salary
distribution, including players where $\hat{\beta}_j \ne 0$.  With the
exception of the bins containing the largest salaries, the absolute
tally of $\hat{\beta}_j \ne 0$ players is fairly uniform.  It is
somewhat surprising that a relatively large proportion of top-dollar
players find themselves with player effects of zero.

\section{Full Posterior Estimation and Decision Making}
\label{sec:decision}

The full posterior distribution from our penalized logistic regression
model was estimated via Markov-Chain Monte Carlo simulation, with
details given in Appendix~\ref{sec:methods}.   We will first use the
full posterior distribution to re-examine our player effects $\beta_j$
while accounting for possible covariance between players.  We will then consider 
additional salary information to explore optimal line combinations
and matchups under budget constraints in Section~\ref{sec:matchups}.  Neither of these analyses is possible without samples from the full posterior distribution.

\subsection{Posterior Analysis of Team--player Model}

In contrast with the MAP estimates from Section~\ref{sec:pointe}, samples from the full posterior distribution do not easily
emit a variable selection.  However, these posterior samples do contain much richer information
about relative player ability via the covariance structure of the
$\bs{\beta}$.  One way of analyzing this information is by
constructing a matrix with the posterior probability that each player
is better than every other player.  Specifically, for each player pair
$(i,j)$, we calculate the proportion of samples where $\beta_i^{(t)}$ is larger than $\beta_j^{(t)}$.  

\begin{figure}[ht!]
\centering
\includegraphics[scale=0.48,trim=50 29 35 20]{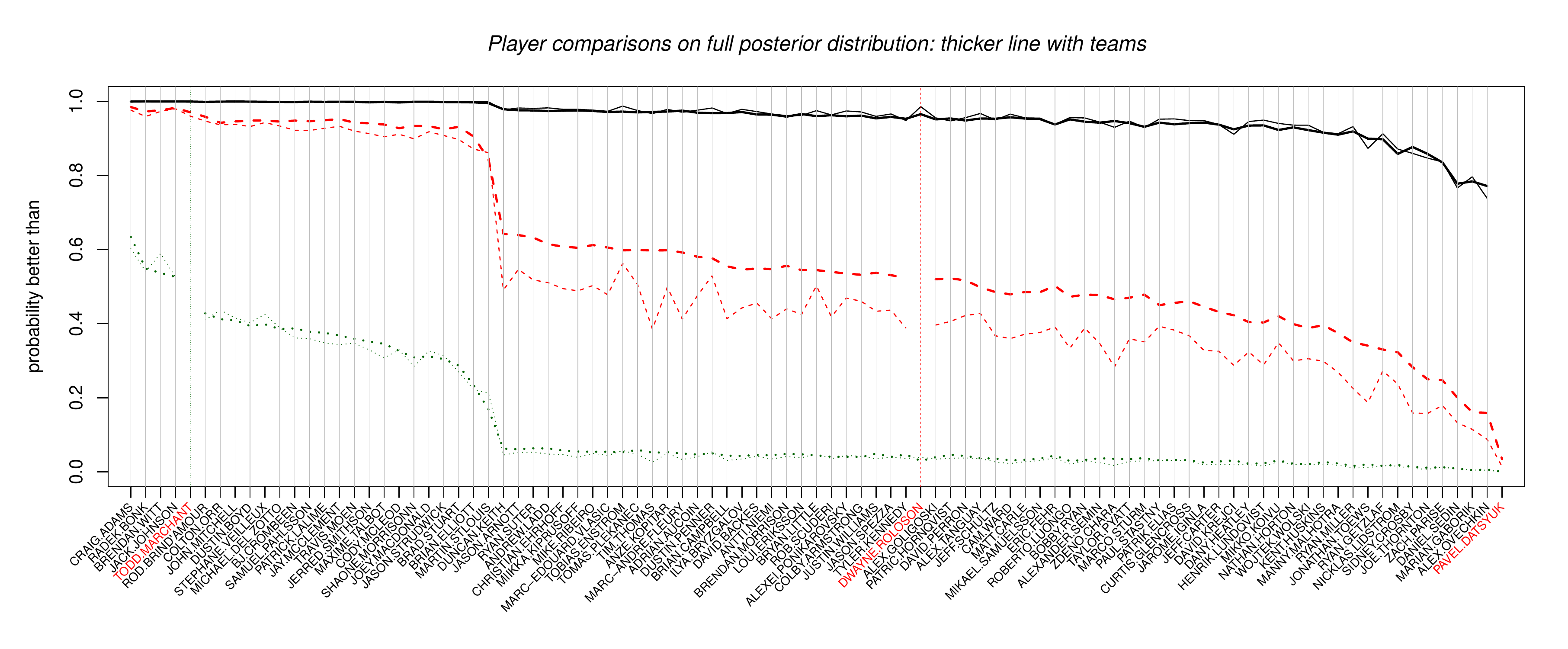}
\caption{Comparing the ability of Datsyuk (black), Roloson (red), and
  Marchant (green) to the 90-odd other players with non-zero
  coefficients in either the team--player or player-only models.
  These three players are also indicated in red among the list of
  players on the X-axis.  Thicker lines correspond to the team--player
  model.}
\label{f:better}
\end{figure}

As an example, Figure \ref{f:better} shows three players (Datsyuk,
Roloson, and Marchant), pitting each of them against the 90-odd
players with non-zero MAP estimates under the team--player model.
Comparisons under both the team--player and player-only models are
provided. Note that the $x$-axis has been slightly re-ordered by the
posterior mean in the team--player model to make the corresponding
curves smoother.

Observe how Roloson's curves indicate a large discrepancy under the
team-player and player-only models (since he played well with poor
teams), whereas Datsyuk's and Marchant's show negligible differences.
We are not aware of any other player effect that can be examined
(pairwise or otherwise) at this resolution, and on such
readily-interpretable probabilistic terms.

\subsection{Posterior Player Match-ups and Line Optimization}
\label{sec:matchups}

Beyond examining pairwise differences,  another way to explore our full posterior
results is through match-ups and line combinations based on the
posterior predictive distribution that accounts for covariances among
our set of players.  Later, we will also build in external constraints
in the form of a cap on salaries.

Specifically, we will calculate the posterior probability that one
particular configuration of players (line A) is more likely to score
or be scored upon when facing another configuration (line B).  This
type of calculation would allow coaches to explore specific line
match-ups against opponents.  In these match-ups, team information
will be ignored but position information respected: we construct only
six-on-six match-ups with one goalie, center, left-wing, right-wing,
and two defensemen on each side.

Consider the following four analyses, where we use our posterior
results to either: 1. pit the best players against the worst players,
2. pit the best players against random players, 3. pit random players
against the worst players, and 4. pit random players against other
random players.  When pitting the best against the worst we construct
the input $x^{(t)}$ for each sample $\bs{\beta}^{(t)}$ as follows.
Place $1$'s in slots for players in each position with the largest
$\bs{\beta}^{(t)}$ value, and $-1$'s in those with the smallest
(largest negative) value.  Fill the rest of the components with zeros.
Then, a sample from the probability that the best team (regarded as
the ``offense") scores is calculated by ${\rm P} ({\rm ``offense"} \, {\rm scores}) =
\mathrm{logit}(x^{(t)\top} \bs{\beta}^{(t)})$ . A comparison of best versus random, worst
versus random or random versus random would proceed similarly
where the random sampling is without replacement.

\begin{figure}[ht!]
\centering
\includegraphics[scale=0.8,trim=10 10 0 40]{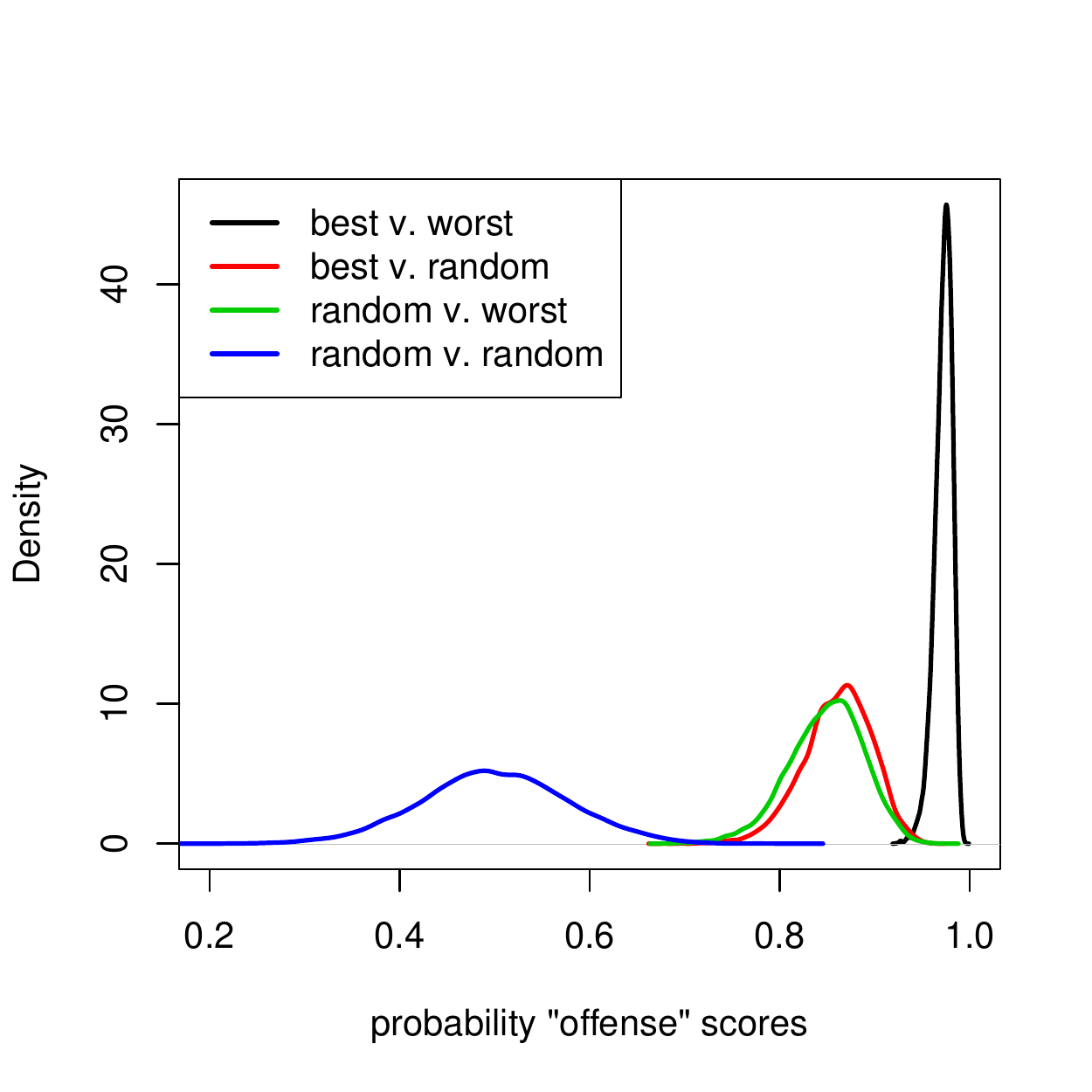}
\caption{Posterior probability that ``offense" scores in various line
  matchups (smoothed using a kernel density).  Better team (listed
  first) is always considered to be the offense. }
\label{f:matchups}
\end{figure}

Figure \ref{f:matchups} shows the results of these matchups, where the
distribution of posterior probabilities that the offense scores are
smoothed using a kernel density.  It is reassuring to see that offense
lines consisting of the best players have a very high probability of
scoring on the worst players and with very low variance.  This is
indicative of a strong signal distinguishing good from bad players in
the data.  Likewise, it is not surprising that the random players
outscore their random counterparts about half the time, and with high
uncertainty.  What is interesting is that the ``best v. random'' and
``random v. worst'' densities are not the same: there is a small but
clear indication in the posterior distribution that the worst players
hurt more than the best players help.

An extended analysis incorporating salary information paints are more
provocative picture.  We now construct our line match-ups subject to a
salary budget/cap $B$, by solving the following binary program:
\begin{align}
\max_{x \in \{0,1\}^{n_p}} &x^\top \bs{\beta}^{(t)}, \nonumber \\
\mbox{subject to} \;\;\; x^\top s &\leq B \label{eq:binprog} \\
\mbox{and} \;\;\;\; x^\top g & = 1, \;\; x^\top \ell  = 1, \;\; x^\top r
= 1, \;\; x^\top d = 2, \nonumber
\end{align}  
where $s$ is an $n_p$-vector of player salaries, and $(g,c,l,r,d)$ are
$n_p$-vectors of binary indicators for goalies, centers, wingers and
defensemen, respectively, so that player positions are still
respected.  The argument of each solution $x^{(t)}$ obtained from the
binary program is then mapped to a posterior sample of the player
effects $\beta^{(t)}$, which gives us the posterior probability
$\mathrm{logit}(x^{(t)\top} \beta^{(t)})$ that the line scores against
a random opponent.

\begin{figure}[ht!]
\centering
\includegraphics[scale=0.5, trim=10 10 0 30]{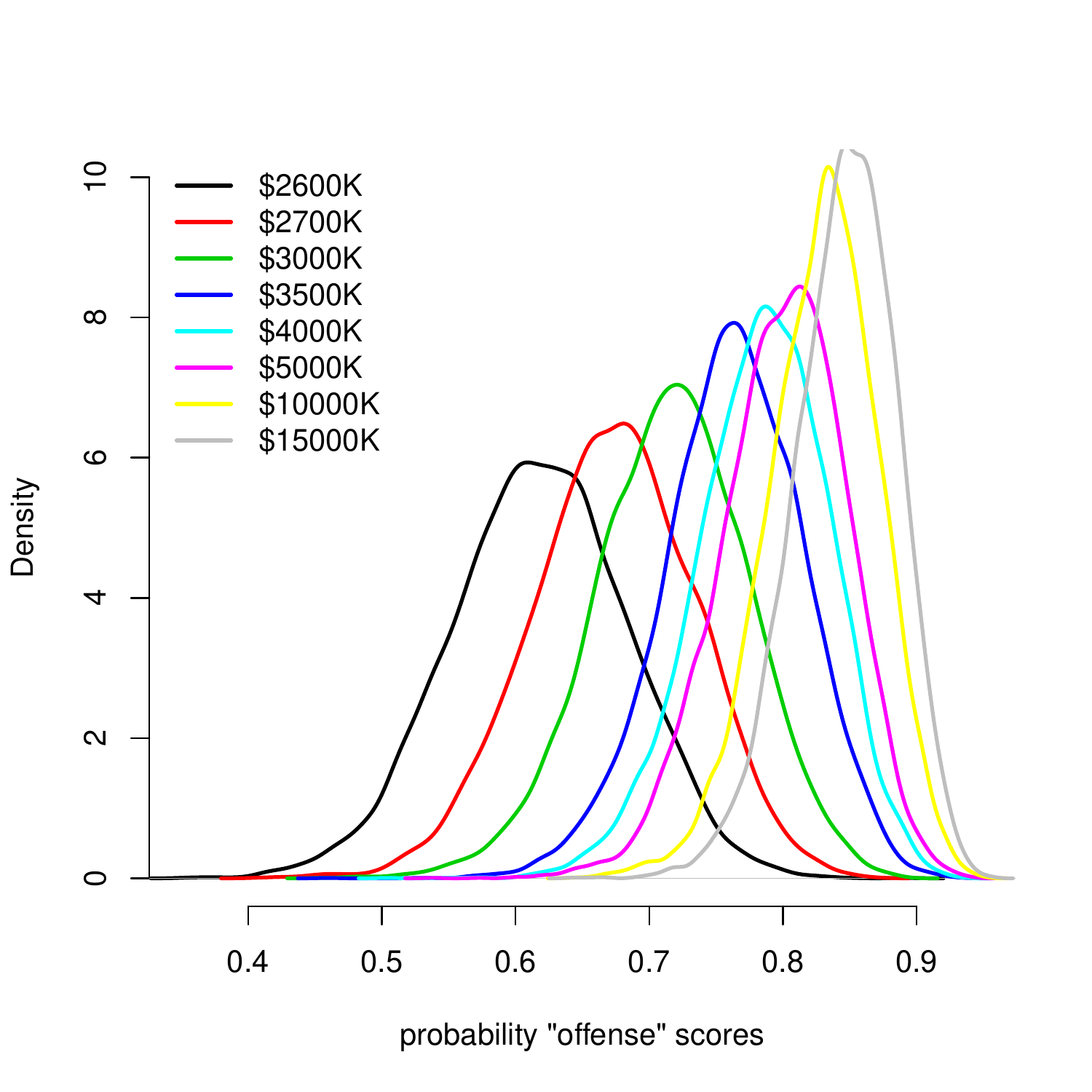} \hspace{0.5cm}
\includegraphics[scale=0.5, trim=10 10 0 30]{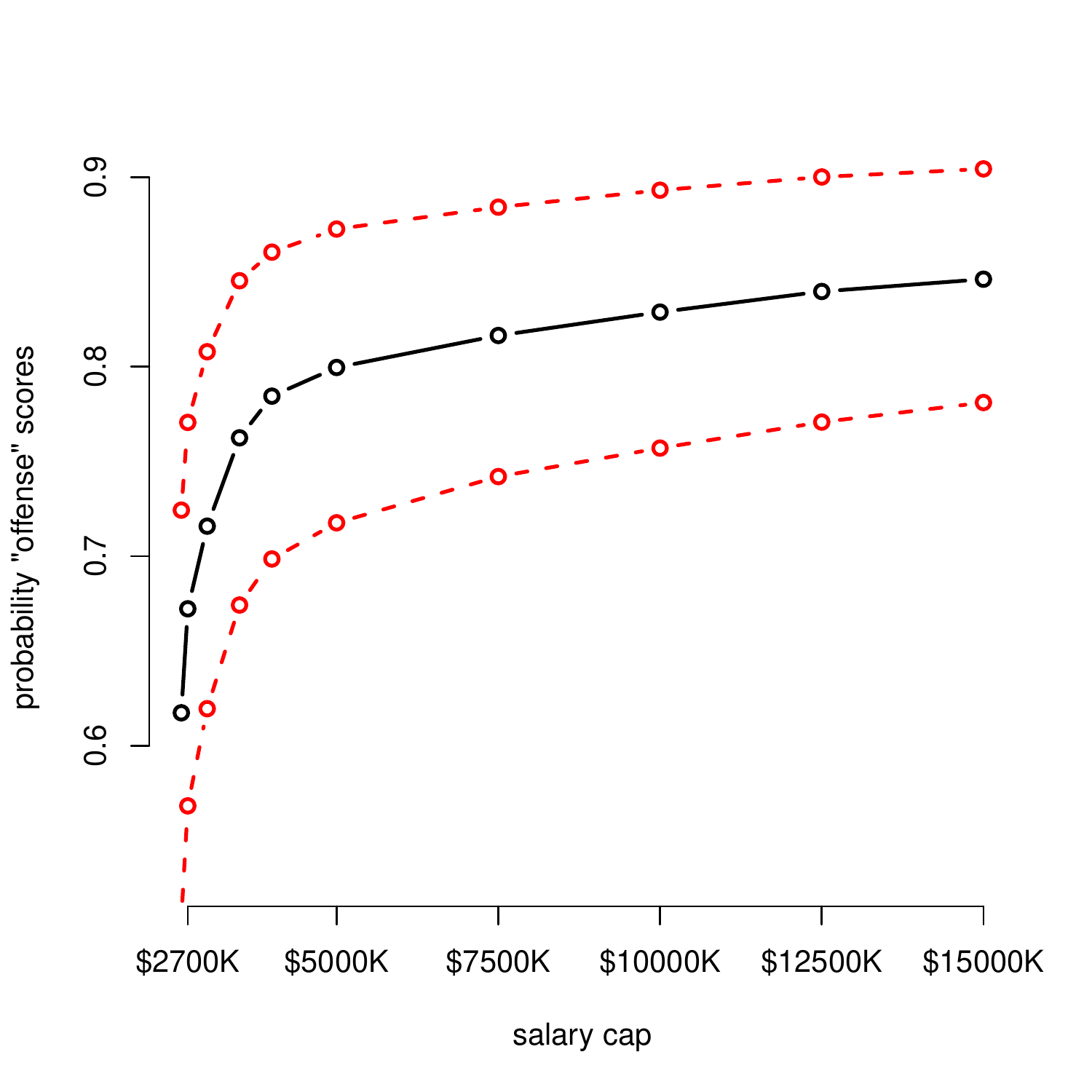}
\caption{The {\em left} panel shows kernel density plots of the
  probability that an optimally chosen line scores against a random
  line according to the full posterior distribution of $\beta$ and
  under several salary caps; the {\em right} panel shows the means and
  90\% predictive intervals of the same posterior as a function of
  those caps.}
\label{f:matchsal}
\end{figure}

The {\em left} panel of Figure \ref{f:matchsal} shows the distribution
of the probability the offense scores for several values of $B$
spanning from \$2.6 million (0.26 $\times$ the current maximum salary
of \$10 million)\footnote{This is the lowest possible budget from
  which lines can be formed satisfying (\ref{eq:binprog}).} to \$15
million.  The {\em right} panel shows the posterior means and 90\%
credible intervals for the probability that the offense scores as a
function of the budget.  The first observation is that there is
tremendous value amongst the cheapest players.  Lines can be formed
among the cheapest players which still outscore their (random)
opponents 65\% of the time, on average.  Importantly, the posterior
probability that this quantity is bigger than 50\% (i.e., that the
best low paid players are better than random ones) is 0.98. A second
observation is that lines formed without the most expensive players,
i.e., with a budget less than (\$10M), are only marginally worse than
those which have these most expensive players.  This means that the
most expensive players may not be good value for money, at least as
far as scoring goals.\footnote{Sweater sales is another matter.}
Having the capacity to have two of the most expensive players on the
ice at once (e.g., Crosby and Malkin for the Penguins) seems to offer
no advantage at all.

Inspecting the individual players that are involved in these optimal
lines in each budget category is also revealing.  Brian Boucher, a
goalie for the Flyers, is extremely valuable for the money, costing
just \$92.5K.  He is the most frequently selected player for each of
the four lowest budgets (\$2.6-3.5M).  Al Montoya (\$750K) is in the
top five of choices in all budgets above the lowest four, representing
a cheap but solid filler player that allows more budget to be
allocated to expensive stars.  At very the top end, Pavel Datsyuk
(\$6.7M) is unsurprising good value for the top three budgets
(\$10-15M).  Ovechkin comes in at a respectable 20$^\mathrm{th}$ place
among all players despite his high cost (\$9M).  Crosby (also \$9M)
comes in the top 25\%.  The most expensive players, Lecavalier and
Luongo, are not selected often at all, suggesting that money is better
spent on cheaper (and younger) talent.


\section{Discussion}
\label{sec:discuss}

In this paper, we use a logistic regression model to estimate the
effects of individual players on goal scoring in hockey.  Unlike the
traditional plus-minus measure that is commonly used, our player 
effects account for the match-ups involved in each goal as well as
overall team contributions.  

We take a Bayesian approach with a prior
distribution that shrinks the coefficients towards zero, which promotes stability in our model and protects against over-fit.  The 
Laplace prior distribution placed on the individual player coefficients allows us to perform variable selection on the large
scale needed for this data situation.  With this variable selection, we are able to separate out a small subset of players
that show substantially above-average performance.

Our analysis gives some
surprising results, such as the dominance of Pavel Datsyuk over other
star players such as Sidney Crosby and Alex Ovechkin.  We also find
that several prominent players, such as Evgeni Malkin, do not have
significant player effects.  

The point estimates $\hat{\bs{\beta}}$ and samples from the full
posterior distribution offer insight into relative player ability at a
resolution not previously available.  Such partial effects and
pairwise comparisons are new metrics derived from making better use of
the same data source behind plus-minus, obtained by leveraging
newfound computational tractability in high dimensional logistic
regression modeling.  We show in our appendix that it is possible to
entertain player--player interaction effects too, although ultimately
conclude that these offer little further insight.

By introducing outside data sources, such as player salary information
and constraints thereon (i.e., salary budgets), our approach offers
unprecedented potential for exploratory analysis, tinkering, and
ultimately decision making by coaches, general managers, and fantasy
players alike.  This analysis of match-ups would only be possible with
our fully Bayesian approach that accounts for the covariance between
individual player effects by using samples from the joint posterior
distribution.

A more believable model for the combination of shift timings and goals
would be to treat every game as a Poisson point process, where each
goal is a point-event and the expected time between goals depends upon
who is on the ice.  However, such an approach requires restrictive
modeling assumptions and considerably more computation, and we doubt
that the value of information about when goals {\it were not} scored
is worth the added complexity.  The popularity of traditional
plus-minus is informative: player ability {\it can} be measured from
the subset of events that actually lead to goals.  Thus while not
completely discounting the potential of time-dependent modeling, we
present the work herein as a robust analysis that is replicable,
extensible, and based upon easily available data.  Similarly, we have
not used power play or short-handed goals in our analysis, but
extending our model to non-even-strength situations is a promising
direction for future work.

\subsection*{Acknowledgments}

Many thanks to Samuel Ventura and Andrew C.~Thomas for supplying a
cleaned version of the {\tt nhl.com} data.  We are grateful to the
editor, an associate editor, and one referee whose comments on an early
version of the manuscript led to many improvements.

\appendix

\section*{Appendix}

\section{Estimation Details and Software}\label{sec:methods}

Although L1 penalty methods and their Bayesian interpretation have
been known for some time, it is only recently that joint
penalty--coefficient posterior computation and simulation methods have
become available for datasets of the size encountered here.  

Probably the most well-known publicly available library for L1 penalty
inference in logistic regression is the {\tt glmnet} package
\citep{fried:hast:tibsh:2009} for {\sf R}.  Conditional on a single
shared value of $\lambda$, this implementation estimates a sparse set
of coefficients.  A convenient wrapper routine called {\tt cv.glmnet}
allows one to chose $\lambda$ by cross-validation (CV).
Unfortunately, CV works poorly in our setting of large, sparse, and
imbalanced $X_P$, where each model fit is relatively expensive and
there is often little overlap between nonzero covariates in the
training and validation sets.  Moreover, when maximizing (rather than
sampling from) the posterior, a single shared $\lambda$ penalty 
leads to over-shrinkage of significant $\beta_j$ as  penalty
choice is dominated by a large number of spurious predictors.  But use
of CV to choose unique $\lambda_j$ for each covariate would imply an
impossibly large search.

Instead, we propose two approaches, both accompanied by publicly
available software in packages for {\sf R}:  joint MAP inference with
\code{textir}, and posterior simulation with \code{reglogit}.

\subsection{Fast variable selection and MAP inference with \code{textir}}

\cite{taddy:2012} proposes a {\it gamma-lasso} framework for MAP
estimation in logistic regression, wherein coefficients and their
independent L1 penalties are inferred under a conjugate gamma
hyperprior.  An efficient coordinate descent algorithm is derived,
including conditions for global convergence, and the resulting
estimation is shown in \cite{taddy:2012} as superior, in both
predictive performance and computation time, to the more common
strategy of CV lasso estimation under a single shared $\lambda$ (as in
{\tt glmnet}).  Results in this paper were all obtained using the
publicly available \code{textir} package for {\sf R} \citep {textir},
which uses the {\tt slam} \citep{slam} package's simple-triplet
matrices to take advantage of design sparsity.

Prior specification in the gamma-lasso attaches independent gamma
$\mathrm{G}(\lambda_j; s,r)$ hyperpriors on each L1 penalty, with
$\mathrm{E}[\lambda_j] = s/r$ and $\mathrm{var}[\lambda] = s/r^2$,
such that, for $j=1\ldots p$,
\begin{equation}\label{eq:glmodel}
  \pi(\beta_j, \lambda_j) = \mathrm{Laplace}(\beta_j;
  \lambda_j)\mathrm{G}(\lambda_j; s,r) =
  \frac{r^s}{2\Gamma(s)}\lambda_{j}^s e^{-\lambda_{j} \left(|\beta_j|+r\right)},~~s,r,\lambda_j>0.
\end{equation}
Laplace priors are often motivated through estimation {\it
  utility}---the prior spike at zero corresponds to a preference for
eliminating regressors from the model in absence of significant
evidence. Our hyperprior is motivated by complementary
considerations: for strong signals and large $|\beta_j|$, expected
$\lambda_j$ shrinks in the joint distribution to reduce estimation
bias.

This leads to the joint negative log posterior {\it minimization}
objective
\begin{equation}\label{eq:pp}
 l(\bs{\alpha},\bs{\beta})= 
  \sum_{i=1}^{n_g} \log \left ( 1 + \exp\left[- y_i (\bm{x}_{Ti}' \bs{\alpha}
      + \bm{x}_{Pi}' \bs{\beta}) \right ] \right ) +
 \frac{1}{2\sigma^2}\sum_{t=1}^{30}
\alpha_t^2 + \sum_{j=1}^{n_p}s\log(1+ |\beta_j|/r),
\end{equation}
where $s$, $r$ $>0$.  We have set $\sigma = 1$ and $r=1/2$ throughout
Section \ref{sec:pointe}. In choosing $s=\mr{E}[\lambda_j]/2$, we focus on the
conditional prior standard deviation, ${\rm SD}(\beta_j) = \sqrt{2}/\lambda_j$,
for the coefficients.  Hence our value of $s=7.5$, for
$\mr{E}[\lambda_j] = 15$, implies expected ${\rm SD}(\beta_j) \approx
0.095$.  To put this in context, $\exp[3\times 0.095] \approx 1.33$,
implying that a single player increasing his team's for-vs-against
odds by $1/3$ is 3 deviations away from the prior mean.

As an illustration of the implementation, the following snippets show
commands to run our main team--player model (see \code{?mnlm} for details).  With 
\verb!X = cbind(XP,XT)! as defined in Section \ref{sec:data}, the list
of 30 ridge and $n_p$ gamma-lasso penalties are specified 
\begin{verbatim}
pen <- c(rep(data.frame(c(0,1)),30),rep(data.frame(c(7.5,.5)),ncol(XP)))
\end{verbatim}
and the model is then fit
\begin{verbatim}
fit <- mnlm(counts=Y, covars=X, penalty=pen, normalize=FALSE)
\end{verbatim}
where \verb!X! is not normalized since this would up-weight players with little ice-time.

\subsection{Full posterior inference via {\tt reglogit}}

Extending a well-known result by \citet{holmes:held:2006},
\citet{gra:pols:2012} showed that three sets of latent variables could
be employed to obtain sample from the full posterior distribution
using a standard Gibbs strategy \citep{gem:gem:1984}.  The full
conditionals required for the Gibbs sampler are given below for the L1
and $\lambda_j = \lambda$ case (note that $\bs{\beta}$ includes
$\bs{\alpha}$ here for notational convenience).
\begin{align*}
  \bs{\beta} | z, \tau^2, \omega, \lambda &\sim
  \mathcal{N}_p(\tilde{\bs{\beta}}, V)
  & \tilde{\bs{\beta}} & = V (y.X)^\top \Omega^{-1} z, \; y.X \equiv \mathrm{diag}(y)X \\
  \lambda |\bs{\beta} &\sim \mathrm{G}\left( a + p, b + \sum_{j=1}^p |
    \beta_j |\right) & V^{-1} &= \lambda^{2} \Sigma^{-1} D_{\tau}^{-1}
  + (y.X)^\top
  \Omega^{-1} (y.X) \\
  \tau_j^{-2} &\sim \mbox{Inv-Gauss} (|\lambda/\beta_j|, \lambda^2),
  & j&=1,\dots, p \equiv \mathrm{ncol}(X) \\
  z_i | \bs{\beta}, \omega_i, y_i, &\sim \mathcal{N}^+ \!\left( y_i
    x_i^\top
    \bs{\beta} , \omega_i \right), & i&=1,\dots,n \equiv \mathrm{length}(y). \\
  \omega_i | y_i, \lambda &\sim \mbox{See below} &
  i&=1,\dots,n
\end{align*}

Note that $\mathcal{N}^+$ indicates the normal distribution truncated
to the positive real line, $D_\tau = \mathrm{diag}(\tau_1^2, \dots,
\tau_p^2)$ and $\Omega = \mathrm{diag}(\omega_1, \dots, \omega_n)$.
\citet{holmes:held:2006} give a rejection sampling algorithm for
$\omega_i|z_i$, however \citet{gra:pols:2012} argue that it is
actually more efficient to draw $\omega_i | y_i, \lambda$ (i.e.,
marginalizing over $z_i$) as follows.  A proposal $\omega_i' =
\sum_{k=1}^K 2 \psi_k^{-1} \epsilon_k$, and $ \epsilon_k \sim
\mathrm{Exp} (1)$ may be accepted with probability $\min\{1,A_i\}$
where $A_i = \Phi\{(- y_i x_i^\top \bs{\beta})/\sqrt{\omega_i'}\}/\Phi\{(-
y_i x_i^\top \bs{\beta})/\sqrt{\omega_i}\}$. Larger $K$ improves the
approximation, although $K=100$ usually suffices.  Everything extends
to the L2 case upon fixing $\tau_j^2 = 1$.  Extending to separate
$\lambda_j$ is future work.

We use the implementation of this Gibbs sampler provided in the {\tt
  reglogit} package \citep{reglogit} for {\sf R}.  For the $\lambda$
prior we use the package defaults of $a=2$ and $b=0.1$ which are
shared amongst several other fully Bayesian samplers for the ordinary
linear regression context (e.g., {\tt blasso} in the {\tt monomvn}
package \citep{monomvn}).  
Usage is similar to that described for {\tt mnlm}.  To obtain {\tt T}
samples from the posterior, simply call:
\begin{verbatim}
bfit <- reglogit(T=T, y=Y, X=X, normalize=FALSE)
\end{verbatim}
Estimating the full posterior distribution for $\bs{\beta}$ allows
posterior means to be calculated, as well as component-wise variances
and correlations between $\beta_j$'s.  However, unlike MAP estimates
$\hat{\bs{\beta}}$, none of the samples or the overall mean is sparse.
\cite{gra:pols:2012} show how their algorithm can be extended to
calculate the MAP via simulation, but this only provides sparse
estimators in the limit.

Another option is to use the posterior mean for $\lambda$ obtained by
Gibbs sampling on the entire covariate set, and use it as a guide in
MAP estimation.  Indeed, this is what is done in Section
\ref{sec:pointe}, where mean of our shared $\lambda$ from Gibbs
sampling was considered when setting setting priors for
$\mr{E}[\lambda_j]$.

\section{Extension to player--player interactions}
\label{interactionresults}

We explore the possibility of on-ice synergies or mismatches between
players by adding interaction terms into our model.  The extension is
easy to write down: simply add columns to the design matrix that are
row-wise products of unique pairs of columns in the original $X_P$.
However, this implies substantial computational cost: see
\cite{gra:pols:2012} for examples of regularized logistic regression
estimators that work well for estimating main effects but break down
in the presence of interaction terms.

There is a further representational problem in the context of our
hockey application.  There are about 27K unique player pairs observed
in the data.  Combining these interaction terms with the original
$n_p$ players, $n_g$ goals and thirty teams, we have a
double-precision data storage requirement of nearly a gigabyte.  While
this is not a storage issue for modern desktops, it does lead to a
computational bottleneck since temporary space required for the linear
algebra routines cannot fit in fast memory.  Fortunately, our original
design matrix has a high degree of sparsity, which means that the
interaction-expanded design matrix is even more sparse, and the sparse
capabilities of \code{textir} makes computation feasible.  Inference
on the fully interaction-expanded design takes about 20 seconds on an
Apple Mac Desktop.
All other methods we tried, including \code{reglogit}, failed in
initialization stages.

\begin{figure}[ht]
\centering
\includegraphics[width=6.75in]{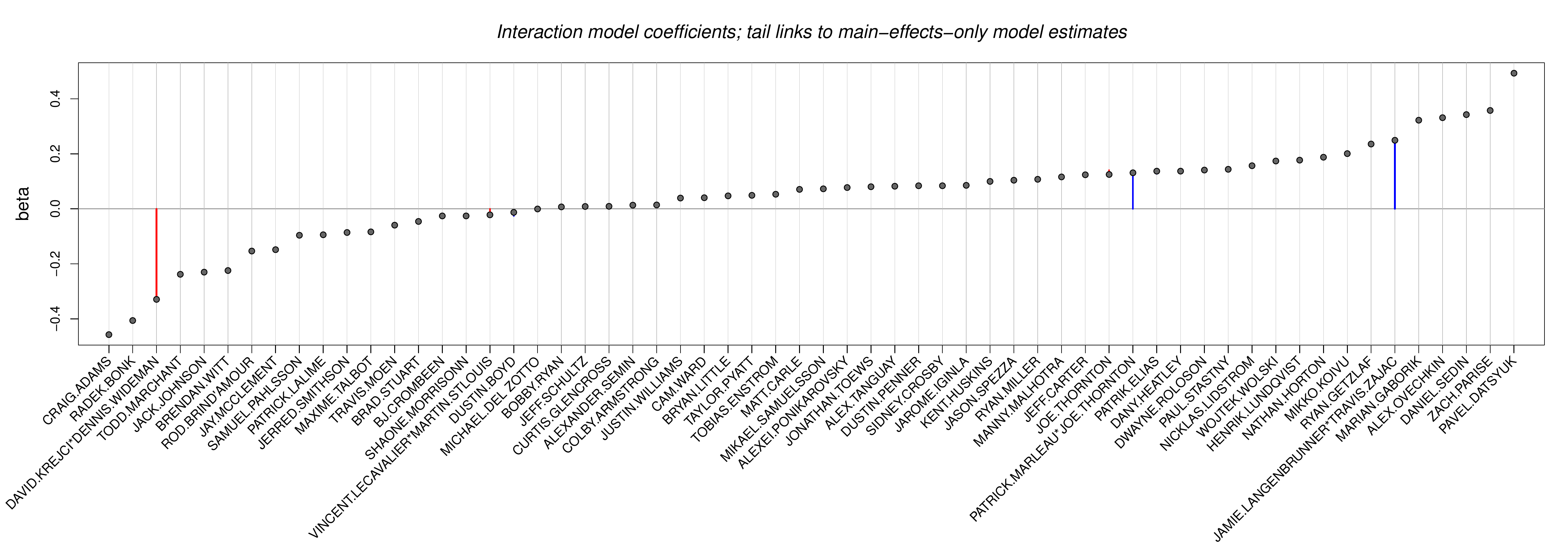}
\caption{Comparing (non-zero) main effects for team--player model 
  to their values in the interaction-expanded team--player model (dots)
  The lines point to the unexpanded estimates, and the $x$-axis is
  ordered by the dots.}
\label{f:maini}
\end{figure}

While the computational feat is impressive, our results indicate
little evidence of significant player interaction effects.  Figure
\ref{f:maini} shows results for estimation with $\mr{E}[\lambda =
15]$: only four non-zero interactions are found when augmenting the
team--player model to include player--player interaction
terms.\footnote{We omitted goalie--skater and goalie--goalie
   interaction terms.}  
 Importantly, there is a negligible effect of including these
 interactions on the individual player effect estimates (which are our
 primary interest).  The only player with a visually detectable
 connecting line is Joe Thornton, and to see it you may need a
 magnifying glass.  We guess from the neighboring interaction term
 that his ability is enhanced when Patrick Marleau is on the ice.  The
 most interesting result from this analysis involves the pairing of
 David Krejci and Dennis Wideman, which has a large negative
 interaction.  One could caution Bruins coach Claude Julien against
 having this pair of players on the ice at the same time.

Using {\tt reglogit} to obtain samples from the full posterior
distribution of the interaction-expanded model is not feasible, but we
can still glean some insight by sampling from the posterior
distribution for the simpler model with the original team-player
design augmented to include the four significant interactions found
from our initial MAP analysis.
\begin{figure}[ht!]
\centering
\includegraphics[scale=0.48,trim=50 29 35 10]{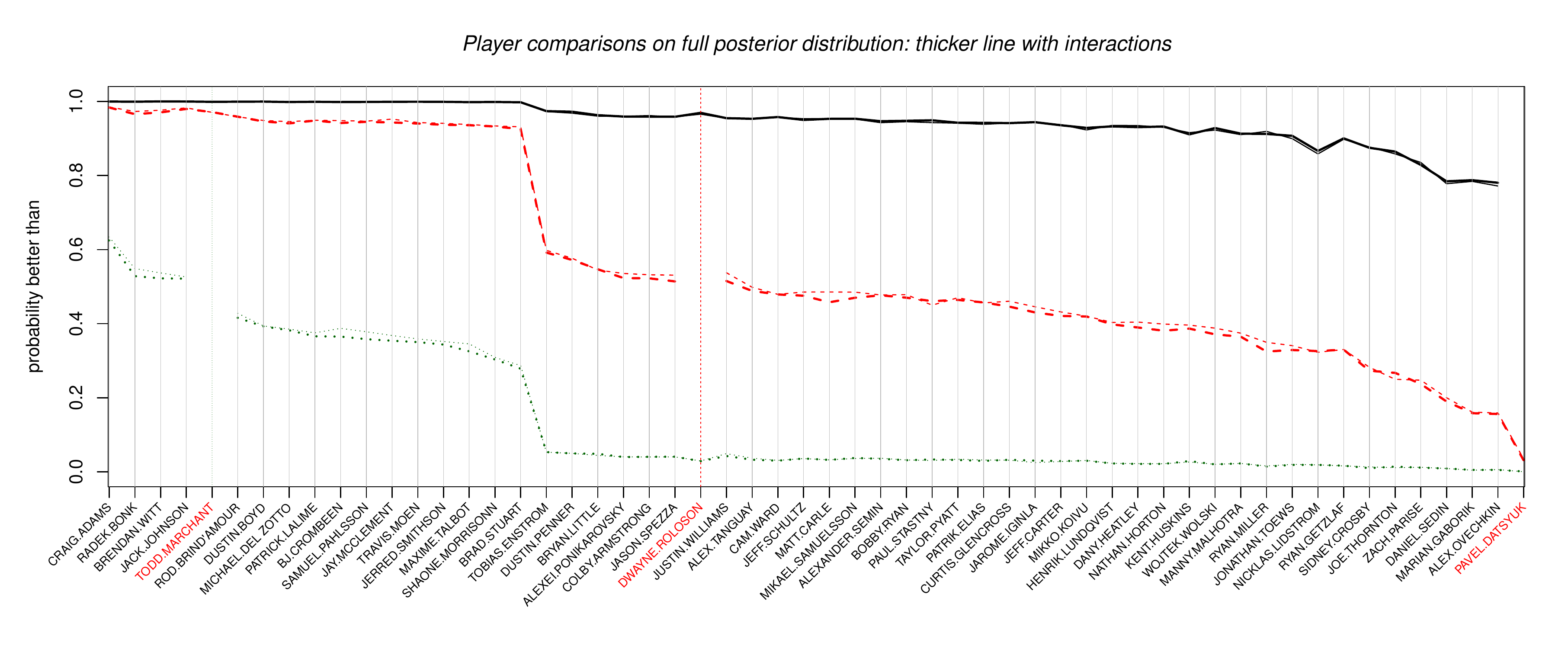}
\caption{Comparing the ability of Datsyuk, Roloson, and Marchant to
  the 60-odd other players with non-zero coefficients in the
  team--player model, showing coefficients under the
  interaction-expanded model as well.}
\label{f:betteri}
\end{figure}
Figure \ref{f:betteri} compares pairwise abilities for the same three
players used as examples in Figure \ref{f:better}.  We observe minimal
changes to the estimates obtained under the original team--player
design, echoing the results from the MAP analysis.  In total, we
regard the ability to entertain interactions as an attractive feature
of our methodology, but it does not change how we view the relative
abilities of players.

\bibliography{pmlogit}
\bibliographystyle{jasa}

\end{document}